\documentclass[a4paper]{article}
\begin{document}
\topmargin -0.2cm \oddsidemargin -0.2cm \evensidemargin -1cm
\textheight 22cm \textwidth 12cm

\title{ Light Bosons of Electromagnetic Field  and  Breakdown of Relativistic Theory.}
  
\author {Minasyan V.N.}
\maketitle

\begin{abstract}  
In our analysis, a quantisation scheme for local electromagnetic waves in vacuum  
is introduced by the model of nonideal Bose-gas consisting of Bose-light-particles 
(which are no photons) with spin one and a finite mass. This fact destroys the Relativistic Theory of Einstein as well as displays a wrong sound of so-called "spontaneous breakdown of symmetry" because the light boson can be moved by speed of wave in vacuum. 
\end{abstract}
PACS:$01.55.+b$ General physics                                                                                                                                                   

\vspace{137mm}

\vspace{5mm}

{\bf 1. INTRODUCTION.} 

\vspace{5mm}

The field of prediction elementary particles is a very complex, 
and as yet not fully solved, Problem. The study of field theoretical 
models connected with spontaneous breakdown of symmetry (there is 
presence a non-zero vacuum expectation value) which occurs, 
if Lagrangian fully invariant under an internal Lee group were 
proposed by different authors [1-4]. Following initial works, done by the Freund and Namby [2], Goldstone [5] and Higgs [6], exhibit on display the bosons as yields of Lorentz invariant relativistic field theory. In this letter, we predict an existence of light-bosons of electromagnetic field. This reasoning destroys a concept of the non-zero vacuum, and in turn the existence of Freund and Namby scalar bosons, Goldstone massless bosons and Higgs massive bosons because predicted light-bosons of electromagnetic field can be moved by speed of electromagnetic wave in vacuum.
 
Theoretical description of quantization local electromagnetic 
field in vacuum within a model of Bose-gas of local electromagnetic waves, which are propagated by speed  $c$ in vacuum, was first proposed by Dirac [7]. 

\vspace{5mm}
{\bf 11. DIRAC THEORY}
\vspace{5mm}

For beginning, we present the Maxwell equations in the zero-vacuum:

\begin{equation}
curl {\vec {H}} -\frac{1}{c}\frac{d {\vec{E}}}{d t}=0
\end{equation}

\begin{equation}
curl {\vec {E}} +\frac{1}{c}\frac{d {\vec{H}}}{d t}=0
\end{equation}

\begin{equation}
div {\vec {E}} =0
\end{equation}

\begin{equation}
div {\vec {H}} =0
\end{equation}

where  $\vec {E}=\vec {E}(\vec {r},t)$ and $\vec {H}=\vec {H}(\vec {r},t)$ are, respectively, the local electric and magnetic fields presented in dependence of the coordinate $\vec {r}$ and current time $t$; $c$ is the velocity of wave in vacuum. 

The Hamiltonian of radiation $\hat{H}_R$ is determined as: 

\begin{equation} 
\hat{H}_R =\frac{1}{8\pi V}\int \biggl( E^2+H^2\biggl) dV 
\end{equation}

In this respect, the Dirac proposed to examine a quantization scheme of electromagnetic field by introducing of the vector potential for local electromagnetic field $\vec {A}(\vec{r},t)$:

\begin{equation} 
\vec {H}= curl  {\vec {A}}
\end{equation} 
and
\begin{equation} 
\vec {E}=- \frac{1}{c}\frac{d {\vec{A}}}{d t}
\end{equation} 
  
which by inserting in (1)-(4), determines a wave-equation:

\begin{equation} 
\nabla^2 {\vec {A} }-\frac{1}{c^2}\frac{d^2 \vec{A}}{d t^2}=0 
\end{equation} 
with condition
\begin{equation} 
div {\vec {A} }=0
\end{equation} 
where

\begin{eqnarray} 
\vec {A}(\vec{r},t)& =&\int  \biggl(\vec {A} _{\vec{k}} \exp^{i(
\vec{k}\vec{r} + kc t )} +\vec {A}^{+}_{\vec{k}}
\exp^{-i(\vec{k}\vec{r} + kc t)} \biggl) d^3 k=\nonumber\\
&=&\sum_{\vec{k}}\biggl(
\vec {A} _{\vec{k}} \exp^{i(
\vec{k}\vec{r} + kc t )} +\vec {A}^{+}_{\vec{k}}
\exp^{-i(\vec{k}\vec{r} + kc t)}\biggl)
\end{eqnarray} 

where $\vec {A} ^{+}_{\vec{k}}$  and $\vec {A} _{\vec{k}}$ are, respectively, the Fourier components of vector potentials electromagnetic field which are considered as the vector Bose-operators "creation" and "annihilation" of a Bose-plane wave with spin one. 

Obviously, we have an expression for component $ {H}_x$ on coordinate $x$:

\begin{eqnarray}
{H}_x& =&\biggl (curl {\vec {A}}\biggl )_x= \frac{d A_z}{dy}-\frac{d A_y}{dz}=
\nonumber\\
&=&\biggl (i\sum_{\vec{k}}\vec {k}\times \biggl(
\vec {A} _{\vec{k}} \exp^{i(
\vec{k}\vec{r} + kc t )} -\vec {A}^{+}_{\vec{k}}
\exp^{-i(\vec{k}\vec{r} + kc t)}\biggl) \biggl)_x
\end{eqnarray}
Then,

\begin{equation}
\vec {H}=curl {\vec {A}}=i\sum_{\vec{k}}\vec {k}\times \biggl(
\vec {A} _{\vec{k}} \exp^{i(
\vec{k}\vec{r} + kc t )} -\vec {A}^{+}_{\vec{k}}
\exp^{-i(\vec{k}\vec{r} + kc t)}\biggl)
\end{equation}

To find $H^2$, we use of a supporting formulae from textbook [8] 
$$
\biggl[\vec {a} \times \vec {b}\biggl]\cdot \biggl[\vec {c} \times \vec {d}\biggl]=( \vec {a}\cdot \vec {c})(\vec {b}\cdot \vec {d}) -
(\vec {a}\cdot \vec {d})(\vec {b}\cdot \vec {c})
$$

which leads to following form for $\frac{1}{8\pi V}\int H^2dV $
at application (12) with using of the condition of transverse wave $\vec{k}\cdot \vec {A}_{\vec{k}}=0$ and 

$$
\frac{1}{V}\int e^{i\vec{k}\vec{r}} dV=\delta_{\vec{k}}
$$

In this respect, we posses 
\begin{eqnarray}
\frac{1}{8\pi V}\int H^2dV&=&-\frac{1}{8\pi}\sum_{\vec{k},\vec{k}_1}\delta_{\vec{k} + \vec{k}_1}  
\vec {k}\vec  {k}_1 
\biggl(\vec {A}_{\vec{k}}  -\vec {A}^{+}_{-\vec{k}}\biggl) 
\biggl(\vec {A}_{\vec{k}_1}  -\vec  {A}^{+}_{-\vec{k}_1}\biggl)=
\nonumber\\
&=&  \frac{1}{8\pi }\sum_{\vec{k}} k^2\biggl(\vec {A} _{\vec{k}}  -\vec  {A}^{+}_{-\vec{k}}\biggl)
\biggl(\vec {A} _{-\vec{k}}  -\vec  {A}^{+}_{\vec{k}}\biggl)
\end{eqnarray}

We now calculate the part of the Hamiltonian radiation in (5) 
\begin{equation}
\frac{1}{8\pi V}\int  E^2 dV=\frac{1}{8\pi c^2 V }
\int \biggl(\frac{d {\vec{A}}}{d t}\biggl)^2 dV 
\end{equation}  

At calculation of value $\frac{d {\vec{A}}}{d t}$, we use of a suggestion proposed by Dirac which implies a consideration of current time $t=0$  [1]:

\begin{equation} 
\frac{d {\vec{A}}}{d t}=ic\sum_{\vec{k}}k\biggl(
\vec {A} _{\vec{k}} -\vec {A}^{+}_{\vec{k}}\biggl)
e^{i\vec{k}\vec{r}}
\end{equation}

Inserting value of $\frac{d {\vec{A}}}{d t}$ from Eq.(15) into Eq.(14), we find the Hamiltonian of radiation $\hat{H}_R $ by following form:

\begin{eqnarray} 
\frac{1}{8\pi c^2 V }
\int \biggl(\frac{d {\vec{A}}}{d t}\biggl)^2 dV& =&
-\frac{ 1}{8\pi }\sum_{\vec{k},\vec{k}_1} \delta _{k+ {k}_1} |k| \cdot |k_1|  \biggl(
\vec {A} _{\vec{k}} - \vec  {A}^{+}_{-\vec{k}}\biggl) \biggl(\vec {A}  _{\vec{k}_1}  -\vec  {A}^{+}_{-\vec{k}_1}\biggl) =
\nonumber\\
&=& - \frac{1}{8\pi }\sum_{\vec{k}} k^2\biggl(\vec {A} _{\vec{k}}  -\vec  {A}^{+}_{-\vec{k}}\biggl)
\biggl(\vec {A} _{-\vec{k}}  -\vec  {A}^{+}_{\vec{k}}\biggl)
\end{eqnarray}
Thus, by using of results (13) and (16), we obtain the Dirac Hamiltonian

\begin{equation} 
\hat{H}_R =\frac{1}{8\pi V}\int  E^2 dV +\frac{1}{8\pi V}\int H^2dV =0 
\end{equation} 

which is not able to describe the Plank photon gas. This fact allows us to suggest that it needs to find a new solution of Maxwell equations, which could provide the description of the Plank photon gas.  
 
\vspace{5mm}
{\bf 111. QUANTIZED MAXWELL EQUATIONS }
\vspace{5mm}

To solve a problem connected with a quantization electromagnetic field, we propose the quantized equations of Maxwell.  For beginning, we search the solution of (1)-(4) by following way:  

\begin{equation} 
\vec {E}=- \frac{\alpha}{c}\cdot
\frac{d {\vec{H}_0}}{d t}+\beta\cdot \vec {E}_0
\end{equation} 
and
\begin{equation} 
\vec {H}=\frac{\alpha}{c}\cdot
\frac{d {\vec{E}_0}}{d t} + \beta \vec{H}_0
\end{equation}

where $\alpha $ and $\beta $ are the constants which we obtain in the bellow by using of a physical property of electromagnetic field; $\vec {E}_0=\vec {E}_0(\vec {r},t)$ and $\vec {H}_0=\vec {H}_0 (\vec {r},t)$ are, respectively, determined as the vectors second quantization wave functions for one Bose particle of electromagnetic field with spin one and  mass $m$. In this context, we claim that the vectors of local electric $\vec {E}_0$ and magnetic $\vec {H}_0$ fields, presented by equations (18) and (19), satisfy to the equations of Maxwell in vacuum which here describe the states of the Bose particles:

\begin{equation}
curl {\vec {H}_0} -\frac{1}{c}\frac{d {\vec{E}_0}}{d t}=0
\end{equation}

\begin{equation}
curl {\vec {E}_0} +\frac{1}{c}\frac{d {\vec{H}_0}}{d t}=0
\end{equation}

\begin{equation}
div {\vec {E}_0} =0
\end{equation}

\begin{equation}
div {\vec {H}_0} =0
\end{equation}
In this context, by using of (20), we can rewrite (19) as 
\begin{equation} 
\vec {H}= \frac{\alpha }{c}\frac{d {\vec{E}_0}}{d t} + \beta \vec{H}_0
\end{equation}

By presentation of new terms $E_0$ and $H_0$, the Hamiltonian of radiation $\hat{H}_R$ in (5) takes a following form: 

\begin{eqnarray} 
\hat{H}_R &=&\frac{1}{8\pi V}\int \biggl( E^2+H^2\biggl) dV = \frac{1}{8\pi V}\int \biggl [\biggl(- \frac{\alpha }{c}\frac{d {\vec{H}_0}}{d t}+\nonumber\\
&+&\beta \vec {E}_0\biggl)^2+\biggl (\frac{\alpha }{c}\frac{d {\vec{E}_0}}{d t}+ \beta \vec {H}_0\biggl)^2\biggl] dV =
\hat{H}_e+\hat{H}_h
\end{eqnarray}

where

\begin{equation} 
\hat{H}_e=\frac{1}{8\pi V}\int \biggl [\biggl(\frac{\alpha}{c}
\frac{d {\vec{E}_0}}{d t}\biggl)^2+
\beta^2\vec {E}^2_0\biggl]dV
\end{equation}

\begin{equation} 
\hat{H}_h=\frac{1}{8\pi V}\int \biggl [\biggl(\frac{\alpha}{c}
\frac{d {\vec{H}_0}}{d t}\biggl)^2+
\beta^2\vec {H}^2_0\biggl]dV
\end{equation}

Obviously, the equations (20)-(23) lead to a following wave-equation:

\begin{equation} 
\nabla^2 {\vec {E}_0}-\frac{1}{c^2}\frac{d^2 \vec{E}_0}{d t^2}=0 
\end{equation}
and
\begin{equation} 
\nabla^2 {\vec {H}_0}-\frac{1}{c^2}\frac{d^2 \vec{H}_0}{d t^2}=0 
\end{equation}
which in turn have following solutions:

\begin{equation} 
\vec {E}_0= \frac{1}{V}\sum_{\vec{k}}\biggl(
\vec {E} _{\vec{k}} e^{i(
\vec{k}\vec{r} + kc t )} +\vec {E}^{+}_{\vec{k}}
e^{-i(\vec{k}\vec{r} + kc t)}\biggl)
\end{equation}

\begin{equation} 
\vec {H}_0= \frac{1}{V}\sum_{\vec{k}}\biggl(
\vec {H} _{\vec{k}} e^{i(
\vec{k}\vec{r} + kc t )} +\vec {H}^{+}_{\vec{k}}
e^{-i(\vec{k}\vec{r} + kc t)}\biggl)
\end{equation} 

where  $\vec { E } ^{+}_{\vec{k}}$, $\vec { H } ^{+}_{\vec{k}}$  and  $\vec {E} _{\vec{k}}$, $\vec {H} _{\vec{k}}$ are, respectively, the second quantzation vectors wave functions, which are represented as the vector Bose-operators "creation" and "annihilation" of the Bose-particles of electric and magnetic waves with spin one. 

We now insert a value of $\vec {E}_0$ from (30) into (26), and then: 

\begin{eqnarray} 
\frac{1}{8\pi V}\int \biggl [\biggl(\frac{\alpha}{c}
\frac{d {\vec{E}_0}}{d t}\biggl)^2dV&=&-\frac{\alpha^2}{8\pi}\sum_{\vec{k},\vec{k}_1} 
\delta_{\vec{k}+\vec{k}_1} |k| \cdot |k_1|  \biggl(
\vec {E}_{\vec{k}} - \vec{E}^{+}_{-\vec{k}}\biggl) 
\biggl(\vec{E}_{\vec{k}_1} -\vec{E}^{+}_{-\vec{k}_1}\biggl) =
\nonumber\\
&=&-\frac{\alpha^2}{8\pi}\sum_{\vec{k}}k^2
\biggl(\vec{E}_{\vec{k}} -\vec{E}^{+}_{-\vec{k}}\biggl)
\biggl(\vec{E}_{-\vec{k}} -\vec{E}^{+}_{\vec{k}}\biggl)
\end{eqnarray}

Consequently, within introducing of assumption that the 
term with square wave vector $k^2$ describes the kinetic 
energy of free Bose-particles of electromagnetic field with mass $m$ by definition $\frac{\alpha^2 k^2}{4\pi}=\frac{\hbar^2 k^2}{2m}$,  we find a constant $\alpha = \frac{\hbar \sqrt{2\pi}}{\sqrt{m}}$. Then, we posses:

\begin{eqnarray} 
\frac{1}{8\pi V}\int \biggl [\biggl(\frac{\alpha}{c}
\frac{d {\vec{E}_0}}{d t}\biggl)^2dV &=&\sum_{\vec{k}}\frac{\hbar^2k^2}{2m}\vec {E}^{+}_{\vec{k}}\vec {E}_{\vec{k}}- 
\sum_{\vec{k}}\frac{\hbar^2k^2}{4m}
\biggl (\vec {E}^{+}_{\vec{k}}
\vec {E}^{+}_{-\vec{k}}+   
\vec {E}_{-\vec{k}}\vec {E}_{\vec{k}}\biggl)- \nonumber\\
&-&\sum_{\vec{k}}\frac{\hbar^2k^2}{4m}
\end{eqnarray}

As we see the first term in right pat of (33) represents as the kinetic energy of the Bose gas consisting of the Bose-particles of electromagnetic field but the second term in right pat of (33) describes the term of the interaction between particles. 
In this context,  the part in $\hat{H}_e$ in (26) takes a following form:
\begin{eqnarray} 
\frac{1}{8\pi V}\int 
\beta^2\vec {E}^2_0dV &=&
\frac{\beta^2}{8\pi}\sum_{\vec{k},\vec{k}_1}
\delta(\vec{k}+ \vec{k}_1)\biggl(\vec {E} _{\vec{k}} +
\vec  {E}^{+}_{-\vec{k}}\biggl) 
\biggl(\vec {E}_{\vec{k}_1} +\vec  {E}^{+}_{-\vec{k}_1}\biggl) 
=\nonumber\\
&=&\frac{\beta^2}{8\pi}\sum_{\vec{k}} 
\biggl(\vec {E} _{\vec{k}}+\vec  {E}^{+}_{-\vec{k}}\biggl)
\biggl(\vec {E} _{-\vec{k}} +\vec{E}^{+}_{\vec{k}}\biggl)=
\nonumber\\
&=&\frac{\beta^2}{8\pi}\sum_{\vec{k}} 
\biggl(2\vec  {E}^{+}_{\vec{k}}\vec {E} _{\vec{k}}+
\vec {E} _{\vec{k}}\vec {E} _{-\vec{k}} +\vec{E}^{+}_{-\vec{k}}\vec{E}^{+}_{\vec{k}}\biggl)+ \frac{\beta^2}{8\pi}\sum_{\vec{k}}1
\end{eqnarray}
Consequently, the  operator $\hat{H}_e$ is presented by a following form:

\begin{equation} 
\hat{H}_e=
\sum_{\vec{k}}\biggl (\frac{\hbar^2
k^2}{2m }+
\frac{\beta^2}{4\pi }\biggl )
\vec{E}^{+}_{\vec{k}}\vec{E}_{\vec{k}}+ 
\frac{1}{2V}\sum_{\vec{k}}\hat{U}_{\vec{k}}
\biggl (\vec{E}^{+}_{\vec{k}}
\vec{E}^{+}_{-\vec{k}}+   
\vec{E}_{-\vec{k}}\vec{E}_{\vec{k}}\biggl)
\end{equation}

where $\hat{U}_{\vec{k}}=-\frac{\hbar^2k^2 V}{2m }+
\frac{\beta^2V}{4\pi }$ in the second term in right side of (35) describes the interaction between the Bose-particles. We claim that the inter-particle interaction $\hat{U}_{\vec{k}}$ represents as a repulsive $\hat{U}_{\vec{k}}>0$ in the space of wave vector $\vec{k}$. This assumption leads to the condition for wave numbers  $k\leq k_0=\frac{\beta }{\hbar}\sqrt{\frac{m}{2\pi}}$ where $k_0$ is the boundary maximal wave number which provides that the existence of the interaction energy $\hat{U}_{\vec{r}}$ between two light bosons in the coordinate space (the form of  $\hat{U}_{\vec{r}}$ will be presented in section V): 

\begin{equation}
\hat{U}_{\vec{r}}=\frac{1}{V}\sum_{\vec{k}} \hat{U}_{\vec{k}}\cdot e^{i\vec{k}\vec{r}}
\end{equation}  

Obviously, the sum in (36) diverges, within introducing the concept of a boundary wave number $k_0$ for electromagnetic field. The light bosons with wave number exceeding the boundary wave number $k_0$ do not exist. 

In analogy manner, we can find

\begin{equation} 
\hat{H}_h=
\sum_{\vec{k}}\biggl (\frac{\hbar^2
k^2}{2m }+
\frac{\beta^2}{4\pi }\biggl )
\vec{H}^{+}_{\vec{k}}\vec{H}_{\vec{k}}+ 
\frac{1}{2V}\sum_{\vec{k}}\hat{U}_{\vec{k}}
\biggl (\vec{H}^{+}_{\vec{k}}
\vec{H}^{+}_{-\vec{k}}+   
\vec{H}_{-\vec{k}}\vec{H}_{\vec{k}}\biggl)
\end{equation}

Thus, the Hamiltonian radiation $\hat{H}_R $ is determined for the Bose gas consisting of the Bose-particles with wave numbers $ k\leq k_0$: 
\begin{equation}
\hat{H}_R=\hat{H}_e+ \hat{H}_h
\end{equation}
where $\hat{H}_e$ and $\hat{H}_h$ are presented by formulas (35) and (37).

To evaluate an energy levels of the operators $\hat{H}_R$ in (38)
within diagonal form, we again apply new linear transformation of vector Bose-operator which is a similar to one for scalar Bose-operator presented by Bogoliubov [9]:

\begin{equation}
\vec {E}_{\vec{k}}=\vec {H}_{\vec{k}}=\frac{\vec {h}_{\vec{k}} + 
M_{\vec{k}}\vec {h}^{+}_{-\vec{k}}} {\sqrt{1-M^2_{\vec{k}}}}
\end{equation} 

where $M_{\vec{k}}$ is the real symmetrical functions  
of  a wave vector $\vec{k}$.

which transforms a form of operator Hamiltonian $\hat{H}_R$ by following way:  
\begin{equation}
\hat{H}_R=
2\sum_{ k\leq k_0}\eta_{\vec{k}}\vec {h}^{+}_{\vec{k}} 
\vec {h}_{\vec{k}}
\end{equation}

Hence, we infer that the Bose-operators 
$\vec {h}^{+}_{\vec{k}}$ and 
$\vec {h}_{\vec{k}}$ are, respectively, 
the vector operators "creation" and "annihilation" of free photons with energy 
 
\begin{equation}
\eta_{\vec{k}}=
\sqrt{\biggl (\frac{\hbar^2k^2}{2m}+\frac{\beta^2}{4\pi }\biggl )^2-\biggl (\frac{\hbar^2k^2}{2m}-\frac{\beta^2}{4\pi }\biggl)^2}=\frac{\hbar k\beta }{\sqrt{2m\pi}}=\hbar k c
\end{equation}

where $\vec {h}^{+}_{\vec{k}}\vec {h}_{\vec{k}}$ is the scalar operator of the number photons occupying the wave vector $\vec{k}$; $c$ is the velocity of photon which defines $\beta =c\sqrt{2m\pi}$ because $c=\frac{ \beta }{\sqrt{2m\pi}}$ in (41). In this respect, the maximal wave number equals to $k_0=\frac{\beta }{\hbar}\sqrt{\frac{m}{2\pi}}=\frac{mc}{\hbar}$.   
 
Thus, the quantized Maxwell equations have following forms:

\begin{equation} 
\vec {E}=- \frac{\hbar \sqrt{2\pi}}{c\sqrt{m}}\cdot
\frac{d {\vec{H}_0}}{d t}+ c\sqrt{2m\pi}\cdot \vec {E}_0
\end{equation} 
and
\begin{equation} 
\vec {H}=\frac{\hbar \sqrt{2\pi}}{c\sqrt{m}}\cdot
\frac{d {\vec{E}_0}}{d t} + c\sqrt{2m\pi} \vec{H}_0
\end{equation}

Our investigation showed that the boson of electromagnetic field has a certainly finite mass $m$. To find the later we states that the source of the photon modes are been the chemical elements which may consider as an ion+electron system which are like to the Hydrogen atom. Due to changing of a electron of its energetic level, by going from high level to low one, leads to an appearance of a photon with energy is determined by a distance between energetic states. The ionization energy of the Hydrogen atom $E_I=\frac{m_e e^4}{2\hbar^2}$ (where $m_e$ and $e$ are the mass and charge of electron) defines the energy of the radiated photon by maximal wave-number $k_0$. Therefore, we may suggest that $\frac{m_e e^4}{2\hbar^2}=\hbar k_0 c$ where $ k_0=\frac{mc}{\hbar}$. This fact discovers a new fundamental constant, which represents as a mass of the light boson:

$$
m=\frac{m_e e^4}{2\hbar^2 c^2}=2.4 \cdot 10^{-35} kg
$$    

\vspace{5mm}
{\bf 1V. CONCLUSSION.}
\vspace{5mm}

In conclusion, we can note that four fundamental particles exist in the nature: 1. the electron with mass $m_e=9\cdot 10^{-31} kg $; 2. the proton with mass $m_p=1.6\cdot 10^{-27} kg $; 3. the neutron with $m_n=1.6\cdot 10^{-27} kg $;  4. the light boson  with mass  $ m=2.4 \cdot 10^{-35} kg $.

Now, we present the form of the interaction energy $\hat{U}_{\vec{r}}$ between two light bosons in the coordinate space in (41), at 
$$
\hat{U}_{\vec{k}}=-\frac{\hbar^2k^2 V}{2m }+
\frac{m c^2 V}{2} >0
$$  
Our calculation shows that

\begin{eqnarray}
\hat{U}_{\vec{r}}&=&\frac{1}{V}\sum_{ k\leq k_0} \hat{U}_{\vec{k}}\cdot e^{i\vec{k}\vec{r}}=4\pi \int^{k_0}_{0} k^2\hat{U}_{\vec{k}} \frac{sin (kr)}{kr}dk=
\nonumber\\
&=&\frac{2\pi V m c^2}{r^3}\biggl[sin \biggl (\frac{mc r}{\hbar}\biggl)+ \frac{mc r}{\hbar} - \frac{mc r}{\hbar} cos \biggl (\frac{mc r}{\hbar}\biggl)\biggl]- 
\nonumber\\
&-&\frac{2\pi V \hbar^2}{m r^5}\biggl [\biggl(\frac{3m^2 c^2 r^2}{\hbar^2}-6\biggl) \cdot sin \biggl (\frac{mc r}{\hbar}\biggl)+ \frac{ m^3 c^3 r^3}{\hbar^3} - \frac{6mc r}{\hbar} -
\nonumber\\
&-&\biggl(\frac{ m^3 c^3 r^3}{\hbar^3} - \frac{6mc r}{\hbar} \biggl)\cdot cos \biggl(\frac{mc r}{\hbar}\biggl) \biggl]
\end{eqnarray}

The existence of a boundary wave number $k_0=\frac{mc}{\hbar}$ for electromagnetic field is connected with the characteristic length of the interaction $\hat{U}_{\vec{r}}$ between two light bosons in the coordinate space  that is a minimal distance $d=\frac{\hbar}{mc}=2.6\cdot 10^{-8} m$ between two neighboring light bosons into the electromagnetic field. We may state herein that the total number of light bosons in volume $V$ is determined as 
$$
\frac{V}{N}=\frac{4\pi d^3}{3}
$$

from which $\frac{N}{V}=1.4\cdot 10^{22} m^{-3}$.

The existence of light-boson with speed $v=c$ in clear vacuum confirms a wrong sound of theories, proposed by Namby, Goldstone and Higgs, which are based on the existence of so-called "vacuons" [10]. On other hand, we see that the maximal momentum of light boson equals to $p_0=mc$ which implies that the maximal speed of light boson is a velocity of light $v=c$ in vacuum. As result of the Relativistic Theory the relativistic mass $M$ is presented as
$$
M=\frac{m}{\sqrt{1-\frac{v^2}{c^2}}}
$$ 
which equals to infinity $M=\infty$, at $v=c$. This fact shows a wrong sound of Einstein Theory.

\newpage
\begin{center} 
{\bf References} 
\end{center} 
 
\begin{enumerate} 
%\bibitem{1}
\item
Y.~Namby~, Phys.Rev.. ~{\bf 117},~648~(1960)
%\bibitem{2}
\item
P.G.O.~Freund~ and Y.~Namby~ , Phys.Rev. Letters ~{\bf 13},~221~(1964)
%\bibitem{3}
\item
S.L.~Glashow~ ., Phys.Rev.. ~{\bf 130},~2132~(1962)
%\bibitem{4}
\item
S.~Coleman~ .and S.L.~Glashow~ ., Phys.Rev. . ~{\bf 134},~ B671~(1964)
%\bibitem{5}
\item
J.~Goldstone~ ., Nuovo Cimento ~{\bf 19},~454~(1961)
%\bibitem{6}
\item
P.W.~Higgs~ ., Phys.Rev.Lett. ~{\bf 12},~132~(1964)
%\bibitem{7} 
\item 
P.A.M..~Dirac~, "The Principles of Quantum Mechanics", ~Oxford at the  
Clarendon  press  
(1958), "Lectures on Quantum Mechanics". ~Yeshiva University New York~  
(1964)  
%\bibitem{8}
\item 
A.~Korn~, M. ~Korn ~, "Mathematical Handbook", ~McGraw –Hill Book company  
%\bibitem{9}
\item
N.N.~Bogoliubov~, Jour. of Phys.(USSR), ~{\bf 11},~23~(1947) 
%\bibitem{10}
\item
J.~ Schwinger ~, Phys.Rev.. ~{\bf 104},~1164~(1954)
\end{enumerate} 
\end{document}